\documentclass{PoS}

\def\Dslash{{\cal D}\!\!\!\!\!\!/}
\def\tr{\,\hbox{tr}\,}

\title{Towards models with a unified dynamical mechanism for elementary particle masses}

\ShortTitle{Dynamical mass}

\author{\speaker{Roberto Frezzotti}\\
        Dipartimento di Fisica, Universit\`a di  Roma ``{\it Tor Vergata}'' and INFN, Sezione di Roma Tor Vergata -- Rome, Italy\\
        E-mail: \email{roberto.frezzotti@roma2.infn.it}}

\author{Giancarlo Rossi\\
        Dipartimento di Fisica, Universit\`a di  Roma ``{\it Tor Vergata}'', and 
INFN, Sezione di Roma Tor Vergata, and Centro Fermi, Museo Storico della Fisica --  Rome, Italy\\
        E-mail: \email{rossig@roma2.infn.it}}

\abstract{Numerical evidence for a new dynamical mechanism of elementary particle mass 
generation has been found by lattice simulation in a simple, yet highly non-trivial 
SU(3) gauge model where 
a SU(2) doublet of strongly interacting fermions is coupled to a complex scalar field 
doublet via a Yukawa and a Wilson-like term.
We point out that if, as a next step towards the construction of a realistic 
beyond-the-Standard-Model model, weak interactions are introduced, then also weak 
bosons get a mass by the very same non-perturbative mechanism. In this scenario fermion 
mass hierarchy can be naturally understood owing to the peculiar gauge 
coupling dependence of the non-perturbatively generated masses. Hence, if 
the phenomenological value of the mass of the top quark or the weak bosons has
to be reproduced, the RGI scale of the theory must be much larger than $\Lambda_{QCD}$. 
This feature hints at the existence of new strong interactions and particles 
at a scale $\Lambda_T$ of a few TeV. In such a speculative framework the electroweak 
scale can be derived from the basic scale $\Lambda_T$ and the Higgs boson should 
arise as a bound state in the $WW+ZZ$ channel.}

\FullConference{The 36th Annual International Symposium on Lattice Field Theory - LATTICE2018\\
		22-28 July, 2018\\
		Michigan State University, East Lansing, Michigan, USA.}

\begin{document}


\section{Elementary particle masses from a ``non-perturbative anomaly''}
\vspace*{-0.1cm}

The Standard Model (SM) of elementary particles, in spite of its very
impressive successes, is widely believed to be only an effective low 
energy theory as it neither accounts for dark matter and the quantum aspects 
of gravity nor provides enough CP-violation for baryogenesis. Moreover,
the SM is by construction unable to shed light on the puzzling problems of 
EW scale naturalness~\cite{THOOFT} and fermion mass hierarchy~\cite{FrogNiel}.
Apart from these still open problems, the authors of Ref.~\cite{BHL89} pointed
out that, if some dynamical mechanism involving non-SM interactions gives rise 
to the mass of the known elementary fermions, the same mechanism also yields 
massive $W^{\pm}$, $Z^0$ bosons and a composite Higgs boson in the $W^+W^-$, 
$Z^0Z^0$, and/or $t \bar t $ channel.

Following the remark of Ref.~\cite{BHL89} and with the aim of solving the
general problems of fermion mass hierarchy and EW scale naturalness, as
well as certain problems~\cite{Hasenfratz:1991it} of Ref.~\cite{BHL89}
with a proper definition of the composite Higgs framework,
in Ref.~\cite{Frezzotti:2014wja} a new mechanism for the dynamical generation
of elementary fermion masses was advocated.
This mechanism is conjectured to be at work in non-Abelian gauge models with fermions
and scalars where 1) (as usual) chiral transformations acting
on fermions and scalars are exact symmetries, but 2) (deviating from common assumptions)
purely fermionic chiral symmetries undergo a hard breaking at the UV cutoff scale.
When bare parameters are ``naturally'' tuned so as to minimize fermion chiral breaking,
in the effective Lagrangian (EL)~\cite{ColeWein} no Yukawa term occurs, but operators of
non-perturbative origin that violate fermion chiral symmetries, among which a
fermion mass term, are expected to appear, if the scalar potential is such that the
theory lives in its Nambu--Goldstone (NG) phase. 

In the proceeding contribution~\cite{MGlat2018} to Lattice2018 convincing numerical evidence
was provided that the mass generation mechanism of Ref.~\cite{Frezzotti:2014wja} is 
indeed realized in the simplest (``toy'', yet highly non-trivial) $d=4$ lattice
gauge model where this phenomenon could take place. To comply with the requirements
1) and 2) above the model considered in Ref.~\cite{MGlat2018} contains \\
$\bullet$ an SU(3) gauge field, $A_\mu^c$ ($c=1,2,...,8$),
with bare (renormalized) coupling $g_0$ ($g_S$), \\
$\bullet$ one Dirac fermion doublet, $Q=(u,d)^T$, transforming as a colour triplet under SU(3), \\
$\bullet$ one complex scalar doublet, $\varphi =(\varphi_0 + i\varphi_3, -\varphi_2 + i\varphi_1)^T$, invariant under SU(3).
Adopting the 2$\times$2 matrix notation $\Phi = [\,\varphi\,|\! -i\tau^2\varphi^*]$,
the toy model Lagrangian, ${\cal L}_{\rm{toy}}(Q,A,\Phi)$, takes the form
\begin{equation}
\hspace{-.06cm}
{\cal L}_{\rm{toy}} \! =\! {\cal L}_{k}(Q,A,\Phi)\!+\!{\cal V}(\Phi)\!+\!{\cal L}_{Wil}(Q,A,\Phi)\!+\! {\cal L}_{Yuk}(Q,\Phi) \, ,\label{TOYLAG}
\end{equation}
with ${\cal L}_{k}$ and ${\cal V}$ representing standard kinetic
terms and the scalar potential. The model has UV cutoff
$\Lambda_{UV} \sim b^{-1}$ and a renormalization group
invariant (RGI) dynamical scale $\Lambda_S$.
Its Lagrangian includes also a Yukawa term, 
${\cal L}_{Yuk}(Q,\Phi) = \eta 
\big{(} \bar Q_L\Phi Q_R+\bar Q_R \Phi^\dagger Q_L\big{)}$, and a term
${\cal L}_{Wil}(Q,\!A,\!\Phi) \! = \! \frac{b^2}{2}\rho\big{(} \bar Q_L{\overleftarrow{\cal D}}_{\!\mu}\Phi {\cal D}_{\!\mu} Q_R + \bar Q_R \overleftarrow{\cal D}_{\!\mu} \Phi^\dagger {\cal D}_{\!\mu} Q_L\big{)}$.
The latter, being a $\Lambda_{UV}^{-2} \times d=6$~operator,
leaves the model power-counting renormalizable~\cite{Frezzotti:2014wja},
exactly like it happens for the Wilson term in lattice QCD~\cite{WilsonLQCD,KarsSmit},
but induces a hard breaking of the purely fermionic chiral symmetries.
Among other symmetries, the Lagrangian~(\ref{TOYLAG}) is
invariant under the global transformations ($\Omega_{L/R} \in {\mbox{SU}}(2)$)
\begin{eqnarray}
\hspace{-.5cm}&&\chi_L\times \chi_R =  [\tilde\chi_L\times (\Phi\to\Omega_L\Phi)]\times [\tilde\chi_R\times (\Phi\to\Phi\Omega_R^\dagger)] \, ,\label{CHIL}\\
\hspace{-.5cm}&&\tilde\chi_{L/R} : Q_{L/R}\rightarrow\Omega_{L/R} Q_{L/R} \, ,\quad \bar Q_{L/R}\rightarrow \bar Q_{L/R}\Omega_{L/R}^\dagger \, . \label{GTWT}
\end{eqnarray}
A (divergent) fermion mass term $\sim \Lambda_{UV} (\bar Q_L Q_R+ {\rm h.c.})$, being
$\chi_L\times\chi_R$ variant, is not generated.

Although the Lagrangian~(\ref{TOYLAG}) is not invariant under the
purely fermionic chiral transformations $\tilde\chi_L\times\tilde\chi_R$,
a critical value of the Yukawa coupling, $\eta_{cr}$,
exists at which the effective
Yukawa term vanishes~\cite{Frezzotti:2014wja,MGlat2018}. 
In the phase with positive renormalized squared scalar mass, $\hat\mu^2_\phi = 
Z^{-1}_{\Phi^\dagger\Phi}(\mu_0^2 - \mu_{cr}^2) >0$, 
where the $\chi_L\times\chi_R$ symmetry is realized {\it \`a la} Wigner,
at $\eta=\eta_{cr}$ the transformations of $\tilde\chi_L\times\tilde\chi_R$
become approximate (up to corrections O($\Lambda_{UV}^{-2}$)) symmetries.
 
In the phase where $\hat\mu^2_\phi<0$, due to the double-well shape of ${\cal V}(\Phi)$,  
{\mbox{$\langle : \! \Phi^\dagger\Phi \!: \rangle= v^2 1\!\!1\, , v\neq 0$}} and the
$\chi_L\times\chi_R$ symmetry is realized \`a la NG, i.e.\ spontaneously broken 
down to SU(2)$_V$.
Moreover, owing to $v \neq 0$, at $\eta=\eta_{cr}$ the residual 
O($\Lambda_{UV}^{-2}$) and $\tilde\chi_L\times\tilde\chi_R$ violating action 
terms polarize the vacuum that is degenerate as a result of the dynamical 
$\tilde\chi_L\times\tilde\chi_R$ spontaneous breaking ensuing from strong interactions. 
In this situation, at $\eta=\eta_{cr}$ the $d\leq 4$ piece of the EL
is conjectured~\cite{Frezzotti:2014wja} to read 
\begin{equation}
\Gamma_4^{NG}|_{\eta_{cr}} \! =  c_2 \Lambda_S^2 \tr(\partial_\mu U^\dagger\partial_\mu U)  + c_1 \Lambda_S [\bar Q_L U Q_R + {\rm h.c.}]  + 
\tilde c \, \Lambda_S R \tr(\partial_\mu U^\dagger\partial_\mu U)
+ \Gamma_{k} + \hat {\cal V} + {\rm O}(\Lambda_S^2 v^{-2}) \; , \label{L4NG}
\end{equation}
with effective scalar potential $\hat {\cal V}$ and kinetic terms 
$ \Gamma_k= \frac{1}{4}(F F)+ \sum_{X=L,R} \bar Q_X\Dslash Q_X 
+\frac{1}{2}\!{\tr}\![\partial_\mu\Phi^\dagger\partial_\mu\Phi] $.
Here, owing to $v \neq 0$, the effective scalar field is conveniently rewritten
in terms of Goldstone ($\zeta_{1,2,3}$) and massive ($\zeta_0$) scalar fields:
$\Phi = R U\; , \quad R= (v+ \zeta_0)\; , \quad  U = \exp [i v^{-1} \tau^k\zeta_k ]\,$,  
where $U$ is a {\em dimensionless} effective field transforming as 
$U \rightarrow \Omega_L U \Omega_R^\dagger$ under $\chi_L \times \chi_R$. It 
represents the exponential Goldstone boson map and is well defined only if $v\neq 0$. 
In Eq.~(\ref{L4NG}) the term $\propto c_1$ describes non-perturbative (NP) breaking of 
$\tilde\chi_L \times \tilde\chi_R$ and provides an effective mass for the 
fermion fields. In fact, expanding $U$ around the identity, one gets
$ c_1 \Lambda_S [\bar Q_L U Q_R + \bar Q_R U^\dagger Q_L] =c_1 \Lambda_S \bar Q Q [1 
+{\mbox{O}}(\zeta/v)]\,$,
i.e.\ a fermion mass term plus a host of more complicated, non-polynomial
$\bar Q-\zeta_{1,2,3} \,{\mbox{particles}}-Q$ interactions.
The coefficient $c_1$ in Eq.~(\ref{L4NG}) has been argued
in Ref.~\cite{Frezzotti:2014wja} to be an O$(g_{S}^4)$,
odd function of $\rho$, with
its $\lambda_0$ dependence arising only at high loop orders. As for its dependence
on the scalar squared mass, $c_1$ is expected to stay finite in the limit
$-\hat\mu_\phi^2 \gg \Lambda_S^2$ (which is of phenomenological
interest, see Refs.~\cite{Frezzotti:2014wja}) and to be non-zero only for
$\hat\mu_\phi^2 <0$. 

A proper understanding of all the NP terms in the EL expression~(\ref{L4NG})
requires considering the natural extension of $\tilde\chi_L \times \tilde\chi_R$ 
symmetry in the presence of weak interactions~\cite{FR18} -- see Sect.~2. 
The form of $\Gamma_4^{NG}|_{\eta_{cr}}$ implies that the renormalized 
$\tilde\chi_L \times \tilde\chi_R$
Schwinger-Dyson equations (SDE) contain in their r.h.s.\ $\tilde\chi$--violating 
NP terms which, if non-zero, must be RGI, as the l.h.s.\ of the SDE is.
Indeed at $\eta_{cr}$ the $\tilde\chi_L \times \tilde\chi_R$ currents have
(independently of $\hat\mu_\phi^2$) zero anomalous dimension~\cite{Frezzotti:2014wja}.
The full NG phase EL, $\Gamma^{NG} \supset \Gamma_4^{NG}$,
contains of course
an infinite set of local terms of arbitrarily high dimension,
among which the RGI operators of NP origin that violate the approximate
$\tilde\chi_L \times \tilde\chi_R$ symmetry. This phenomenon will be
referred to as a {\em ``NP anomaly''} in the $\tilde\chi$ symmetry restoration. 

\vspace*{-0.4cm}
\section{Toy model with one strong and one weak interaction}

\vspace*{-0.2cm}
We consider here an extension of the toy model~(\ref{TOYLAG})
including strong (vector SU(3): coupling $g_S$) and weak 
(chiral SU(2)$_L$: coupling $g_W$)
gauge interactions plus a minimal Dirac fermion content that is
enough to avoid Witten's global SU(2) anomaly~\cite{Witten:1982fp}:
a SU(3)-triplet field $Q$ and a SU(3)-singlet field $N$, whose
left-handed (right-handed) components are in the fundamental
(trivial) representation of the weak SU(2) gauge group. Its classical
Lagrangian takes thus the form
\vspace*{-0.1cm}
\begin{equation}
\hspace{-.06cm}
{\cal L}_{\rm{toy+W}} \! =\! {\cal L}_{k}(Q,N,A,\Phi ,W)\!+\!{\cal V}(\Phi)\!
+\!{\cal L}_{Wil}(Q,N,A,\Phi ,W)\!+\! {\cal L}_{Yuk}(Q,N,\Phi) \, ,\label{TOY+W}
\end{equation}
where ${\cal V}(\Phi)$ is the potential of the scalar $\Phi$-field, 
${\cal L}_{Yuk}(Q,N,\Phi) = \eta_Q \bar Q_L \Phi Q_R + \eta_N N_L \Phi N_R + {\rm h.c.}$
is a standard Yukawa term, while the kinetic and the $\tilde\chi$-violating $d \geq 6$ 
terms read
\vspace*{-0.1cm}
\begin{eqnarray}
& & {\cal L}_{kin} =  \frac{1}{4}F^{A; a}_{\mu\nu}F^{A; a}_{\mu\nu} + \frac{1}{4}F^{W; i}_{\mu\nu}F^{W; i}_{\mu\nu } + \bar Q_L\gamma_\mu {\cal D}_\mu^{A,W} Q_L +\bar Q_R\gamma_\mu{\cal D}_\mu^{A} \,Q_R + \nonumber \\
& & + \bar N_L\gamma_\mu {\cal D}_\mu^{W} N_L +\bar N_R\gamma_\mu\partial_\mu\,N_R + \frac{1}{2}{\rm tr}\big{[} \Phi^\dagger {\overleftarrow{\cal D}}_\mu^W {\cal D}_\mu^W \Phi\big{]} \; ,  \label{TOY+W-KIN}
\end{eqnarray}
\begin{equation}
{\cal L}_{Wil} = \frac{1}{2\Lambda_{UV}^2} \rho\big{(} \bar Q_L{\overleftarrow{\cal D}}_\mu^{A,W} \Phi {\cal D}_\mu^A Q_R+ \bar Q_R \overleftarrow{\cal D}_\mu^A \Phi^\dagger {\cal D}_\mu^{A,W} Q_L 
+ N_L{\overleftarrow{\cal D}}_\mu^{W} \Phi \partial_\mu N_R+\bar N_R \overleftarrow{\partial}_\mu \Phi^\dagger {\cal D}_\mu^{W} N_L
\big{)} \; . \label{TOY+W-WIL}
\end{equation}
If the $W_\mu^{1,2,3}$ bosons transforms in the adjoint representation of SU(2)$_L$ 
this model enjoys an exact SU(2)$_L \times $SU(3) gauge symmetry, with the covariant
derivatives acting on $f = Q,\, N$ given by e.g.\
$$ {\cal D}^{A,W}_\mu f_L = (\partial_\mu\!-i \delta_{f,Q} g_S \lambda^a A^a_\mu -ig_W\frac{\tau^i}{2}W^i_\mu) f_L \; , \quad
\bar f_L \overleftarrow{\cal D}\,^{A,W}_\mu = \bar f_L (\overleftarrow{\partial}_\mu +i \delta_{f,Q} g_S \lambda^a A^a_\mu +ig_W \frac{\tau^i}{2}W^i_\mu) \; . $$
Besides Lorentz, translation, time-reversal and CP symmetries, the model~(\ref{TOY+W})
enjoys the global SU(2)$_L \times $SU(2)$_R$ invariance ($\Omega_L$ and $\Omega_R$
below are independent SU(2) matrices)  
\begin{equation}
 \chi_L \equiv \tilde\chi_L \otimes \chi_{L}^{\Phi} \; , \qquad
 \chi_R \equiv \tilde\chi_R \otimes \chi_{R}^{\Phi} \; , \label{CHI-LR}
\end{equation}
where $\chi_{L,R}^{\Phi}$ acts only on scalars,
$\; \chi_{L}^{\Phi}: ~\Phi\rightarrow\Omega_L\Phi \; , \quad
\chi_{R}^{\Phi}: ~\Phi\rightarrow \Phi\Omega_R^\dagger \; $,
while $\tilde\chi_R$ acts on right-handed fermions $f_R = Q_R, N_R$ and 
$\tilde\chi_L$ on left-handed fermions $f_L = Q_L, N_L$ and $W_\mu$ bosons
(as necessary for invariance of the $Q_L$ and $N_L$ kinetic terms under $\chi_L$)
\begin{equation}
f_R \rightarrow \Omega_{L} f_R \; , \quad 
\bar f_R \rightarrow \Omega_R^\dagger \bar f_R \; ,
\qquad f \in \{Q,N \} \; , \label{CHITIL-R}
\end{equation}
\begin{equation}
f_L \rightarrow \Omega_{L} f_L \; , \quad 
\bar f_L \rightarrow \Omega_L^\dagger \bar f_L \; , \quad
W_\mu \rightarrow\Omega_{L} W_\mu \Omega_L^\dagger \; ,
\qquad f \in \{Q,N \} \; . \label{CHITIL-L}
\end{equation}
Neither $\tilde \chi_{L,R}$ nor $ \chi_{L,R}^{\Phi}$ transformations
alone are symmetries of the model~(\ref{TOY+W}). For $g_W \neq 0$,
not only ${\cal L}_{Yuk}$ and ${\cal L}_{Wil}$ but also the $Q_L$ and $N_L$
kinetic terms are not left invariant by $\tilde \chi_{L}$ transformations.
%


The study of the bare SDE of $\tilde\chi_L$ transformations shows that, owing to 
the symmetries of the model~(\ref{TOY+W}), the $\tilde\chi_L$-violating operators
arising from ${\cal L}_{Yuk}$ and ${\cal L}_{Wil}$ and the fermion kinetic terms in
${\cal L}_{toy+W}$ mix only with two relevant $d=4$ operators, namely the 
$\tilde\chi_L$-variations of the Yukawa and the $\Phi$-kinetic term (the mixing 
coefficients are denoted below by $\bar\eta_L$ and $-\bar\gamma$). The 
renormalized form of the effective (isotriplet) $\tilde\chi_L$-SDE 
will thus read in synthetic operator notation
\begin{eqnarray}
& & Z_{\tilde J} \partial_\mu \tilde J^{L,\, i}_\mu = (\bar\eta_L -\eta)
\sum_{f=Q,N} ( \bar f_L\frac{\tau^i}{2}\Phi f_R - {\rm h.c.} ) + (1 - \bar\gamma)
\frac{i}{2}g_W \tr\Big{(}\Phi^\dagger [\frac{\tau^i}{2},W_\mu] {\cal D}_\mu^W \Phi 
 + {\rm h.c.} \Big{)} 
+ \; \dots \; , \nonumber   \\
& & \hspace*{-0.0cm}
\tilde J^{L,\, i}_\mu = \sum_{f=Q,N} \Big\{ \bar f_L\gamma_\mu\frac{\tau^i}{2}f_L -\frac{b^2}{2}\rho \Big{(}\bar f_L\frac{\tau^i}{2}\Phi {\cal D}^A_\mu f_R - \bar f_R\overleftarrow {\cal D}\,^A_\mu\Phi^\dagger\frac{\tau^i}{2} f_L\Big{)} \Big\} \, + \, 
i g_W\tr\Big{(}[W_\nu,F^W_{\mu\nu}]\frac{\tau^i}{2}\Big{)} \; ,
\label{SDER-TILCHI-L}
\end{eqnarray}
where we omit irrelevant ${\rm O}(\Lambda_{UV}^{-2})$ terms and ellipses stand 
for possible contributions from effective operators related to ``NP anomalies'', 
involving the field $U= \Phi/\sqrt{\Phi^\dagger \Phi}$ and to be discussed below.  
Eq.~(\ref{SDER-TILCHI-L}) implies that the condition of maximal restoration of 
the $\tilde\chi_L$ symmetry amounts to
\begin{equation}
\eta_{cr} = \bar\eta_L(g_s,g_W;\lambda_0; \eta_{cr},\rho_{cr}) \; , 
\qquad 1 = \bar \gamma(g_s,g_W;\lambda_0; \eta_{cr},\rho_{cr}) \; ,
\label{CRITPAR}
\end{equation}
which is satisfied for suitable critical values, $\eta_{cr}$ and
$\rho_{cr}$ of the bare parameters $\eta$ and $\rho$. Below we shall
refer to the model~(\ref{TOY+W}) at $\eta_{cr}$ and $\rho_{cr}$ as 
to the ``critical model''.

The meaning of the condition of maximal $\tilde\chi_L $ symmetry 
restoration is particularly transparent in the Wigner phase of the model, 
where $\hat\mu_\phi^2 >0$ and ``NP anomalies'' cannot occur because $v=0$ and 
the field $U= \Phi/\sqrt{\Phi^\dagger \Phi}$ entering their effective theory 
description is not defined. In this case the condition~(\ref{CRITPAR}) makes
the $\tilde\chi_L $-SDE~(\ref{SDER-TILCHI-L}) to look as a symmetry Ward identity, 
up to irrevant ${\rm O}(\Lambda_{UV}^{-2})$ terms. 
As for the EL of the critical model,
$\Gamma^{Wig}_{cr}$, this means that its $d \leq 4$ piece reads
\begin{equation}
\Gamma^{Wig}_{4, cr} \equiv \Gamma_{4, cr}^{\hat \mu_\phi^2>0} =
\frac{1}{4}[(F^A F^A) + (F^W F^W)]
+ \sum_{f=Q,N} [\bar f_L \Dslash^{A,W} f_L + \bar f_R \Dslash^{A} f_R] +
{\cal V}_{eff}^{\hat \mu_\phi^2>0}[\Phi]  \; . \label{ELCRIT-WIG}
\vspace*{-0.2cm}
\end{equation}
The absence of an effective Yukawa term in $\Gamma^{Wig}_{4, cr}$ implies 
that $\tilde\chi_R $ symmetry is recovered, too, while the cancellation of 
the effective $\Phi$-kinetic terms means that the elementary scalar field gets 
completely decoupled at all physical momentum scales (i.e.\ well below 
$\Lambda_{UV} \to \infty$).
\begin{figure}[htbp]
\centerline{
\hspace{-0.5cm} \includegraphics[scale=0.300,angle=0]{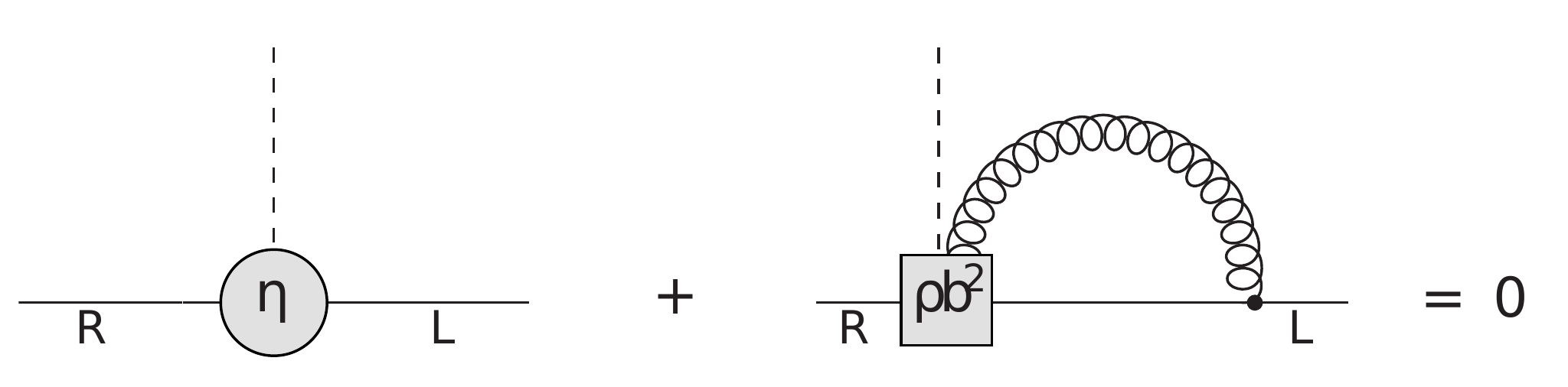}
\hspace{1.5cm} \includegraphics[scale=0.310,angle=0]{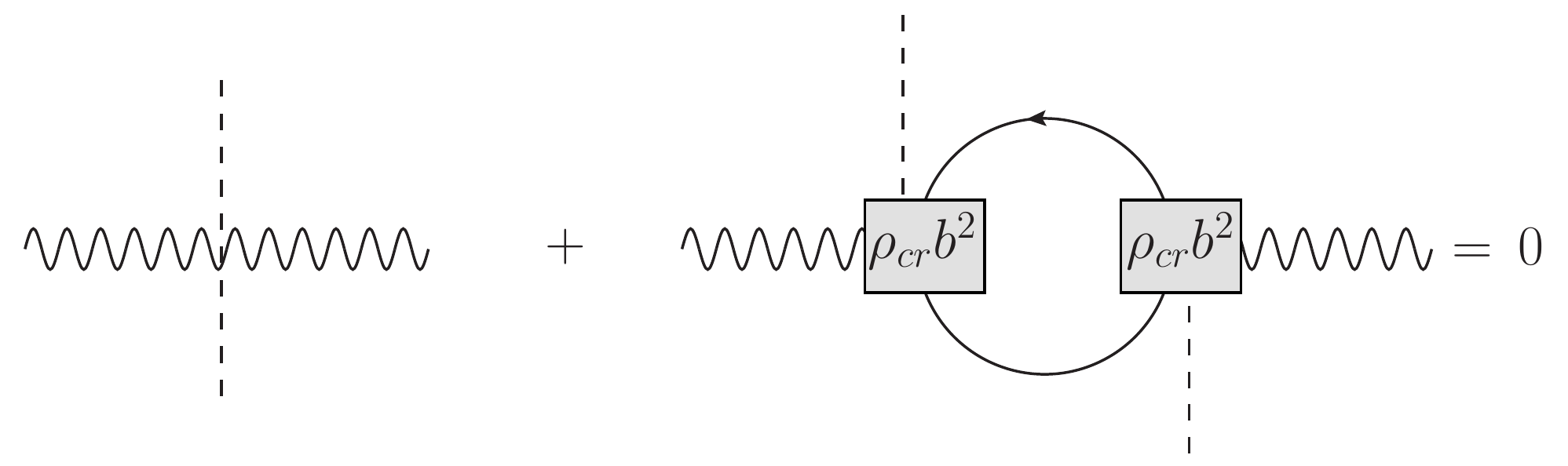}  
}
\caption{The vanishing in the critical model of the effective Yukawa vertex
(left panel) and the effective $\Phi \Phi^\dagger W W$ vertex (right panel)
is illustrated at the lowest non-trivial order of perturbation theory.}
\label{fig:1proc}
\end{figure}
%

In the NG phase, where $\hat\mu_\phi^2 <0$, the EL of the critical model
still has vanishing effective Yukawa and $\Phi$-kinetic terms, but  
includes the terms describing the ``NP anomalies'' that are expected
(known, according to the numerical evidence of Ref.~\cite{MGlat2018} -- see 
Sect.~1) to occur owing to $v >0$ and the fully realized spontaneous breaking 
of the approximate (restored) $\tilde\chi_L \times \tilde\chi_R $ symmetry:
\begin{equation}
\Gamma_{cr}^{NG} \supset \Gamma_{4, cr}^{NG} \; , \qquad 
\Gamma_{4, cr}^{NG} = C_2 \Lambda_S^2 \tr({\cal D}^W_\mu U^\dagger {\cal D}^W_\mu U)  
+ C_{1,Q} \Lambda_S [\bar Q_L U Q_R + {\rm h.c.}]   
 + \Gamma_{4, cr}^{\hat \mu_\phi^2 < 0} \; , \label{L4NG+W}
\vspace*{-0.1cm}
\end{equation}
where $\Gamma_{4, cr}^{\hat \mu_\phi^2 <0}$ is analogous to Eq.~(\ref{ELCRIT-WIG}) 
but with $\hat \mu_\phi^2<0$. At this stage several remarks are in order. \\
{\bf Remark I)} $C_{1,N} = 0$ because $N_R$ fermions are sterile, implying that 
${\cal L}_{\rm toy+W}$ is invariant under the shift symmetry~\cite{Golterman:1989df}
$\; N_R(x) \to N_R(x) + c \; , \quad \bar N_R(x) \to \bar N_R(x) + \bar c \;$,
with $c$, $\bar{c}$ Grassmann-number constants, which in turn 
forbids any mass term for the $N$ fields, including also 
$\Lambda_S[\bar N_L U N_R + {\rm h.c.}]$. \\
{\bf Remark II)} From the form of $\Gamma_{4, cr}^{NG}$ in Eq.~(\ref{L4NG+W}) 
we see that $W_\mu$ and $Q$ fields get a dynamical mass, \\
\vspace*{-0.2cm}
\begin{equation}
\vspace*{-0.1cm}
 (M_W^{eff})^2 = g_W^2 C_2 \Lambda_S^2 \; , \qquad 
m_Q^{eff} = C_{1,Q} \Lambda_S \; ,  \label{W+Q+MASS}
\end{equation}
which can be shown~\cite{FR18} to arise from a common NP mechanism where
both $C_2$ and $C_1$ are $\ll 1$ due to multi-loop suppression. In fact,
assuming the same kind of short distance NP vertex corrections that were
conjectured in Ref.~\cite{Frezzotti:2014wja} (see there Eqs.~(4.12)--(4.14) and 
Fig.~5),
which explain the generation of the NP $Q$-fermion mass $m_Q^{eff}$ 
(see Sect. IV-C and Fig.~6 of Ref.~\cite{Frezzotti:2014wja}), one can
understand/predict the occurrence of the $W_\mu$-boson mass term $\propto C_2 $.
\begin{figure}[htbp]
\vspace*{-0.6cm}
\includegraphics[scale=0.15,angle=0]{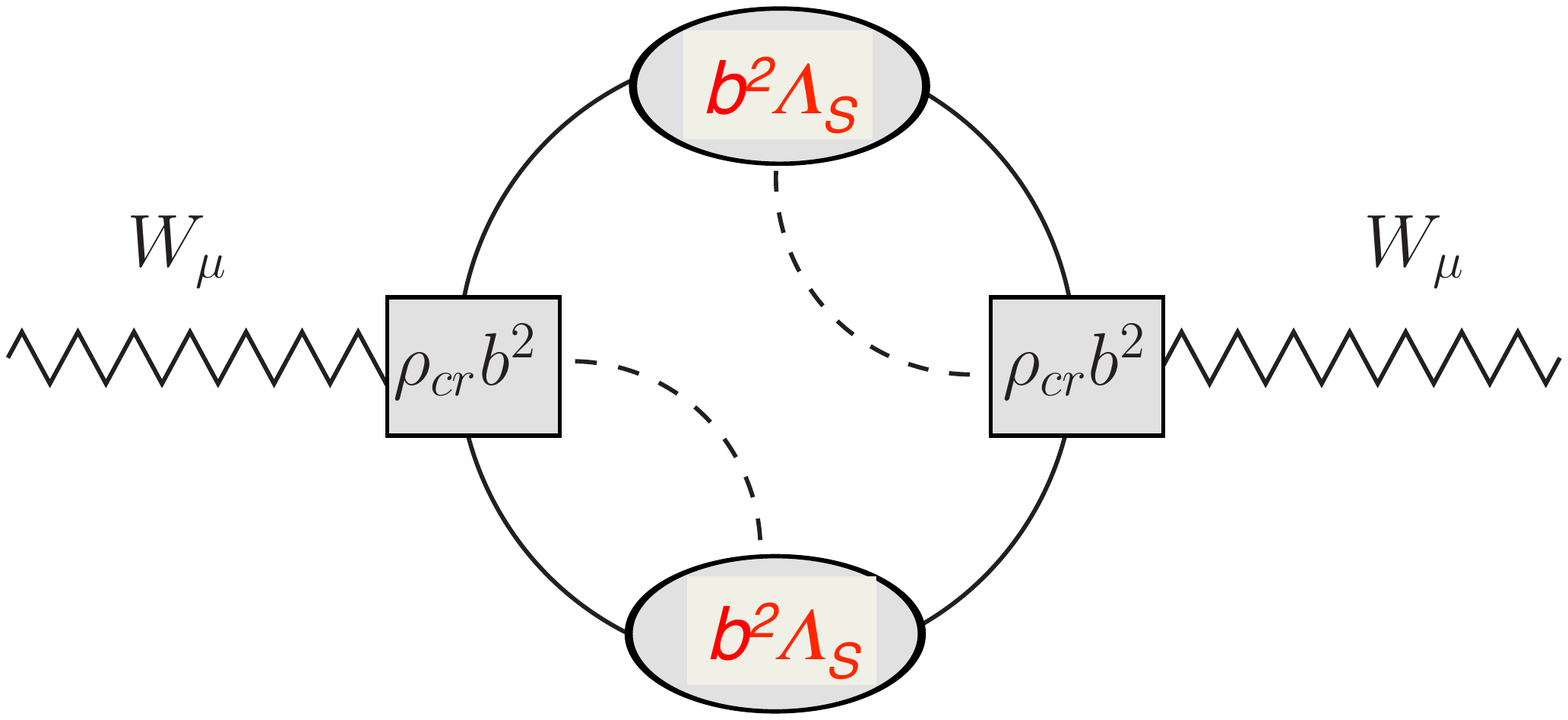} 
\hspace*{4.0cm} 
\includegraphics[scale=0.40,angle=0,trim=0 -70 0 0]{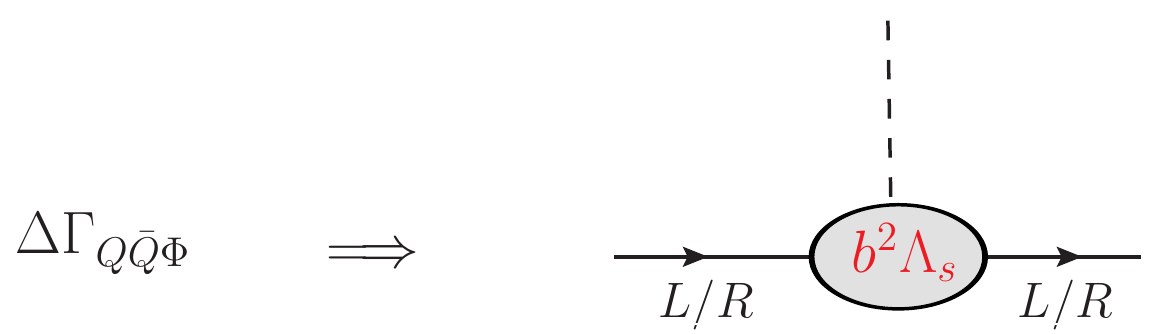}
\vspace*{-0.7cm}
\caption{The lowest loop order ``diagram'' relevant for $(M_W^{eff})^2$ 
(left panel) and the short distance NP vertex $\Delta\Gamma_{Q\bar Q\Phi}$
(right panel). The vertices with a square $\propto b^2\rho_{cr} \equiv
\Lambda_{UV}^{-2} \rho_{cr}$ come from the ${\cal L}_{Wil}$ term of the
Lagrangian~(\ref{TOY+W}). Here and in Fig.~1 lines denote $Q$ fermions, 
wavy lines $W_\mu$ bosons, dashed lines $\Phi$ bosons.} 
\end{figure}
The lowest loop order ``diagram'' relevant for $(M_W^{eff})^2$, which combines 
O($\Lambda_{UV}^{-2}$) $\tilde \chi$ violating vertices and the conjectured short 
distance NP vertices is shown in Fig.~2. \\ 
{\bf Remark III)} In the NG phase of the critical model, owing to the 
``NP anomaly'', the Goldstone boson modes of the elementary $\Phi$  
are still coupled to fermions and $W_\mu$ bosons. The canonically
normalized Goldstone boson fields $\zeta_{1,2,3}$ are hence given by
$U= \Phi /\sqrt{\Phi^\dagger \Phi}= \exp [i\zeta_j\tau^j / \Lambda_S\sqrt{C_2}]$.
From $\Gamma_{4, cr}^{NG}$ one sees that the $WW$-propagator comes out
to be properly transverse owing to the contribution from effective vertices 
with one $W_\mu^j$ and one $\zeta_{j}$ field. In unitary gauge the Goldstone 
bosons are eaten up by the 
massive $W_\mu$ bosons -- just as in the standard Higgs mechanism. \\ 
{\bf Remark IV)} The non-Goldstone mode $\zeta_0$ in $|\Phi| = R = v+\zeta_0$
is instead an auxiliary and fully decoupled field in the critical model at all
physical momentum scales, because of the absence in $\Gamma_{4, cr}^{NG}$ of 
the operator $ \tr({\cal D}^W_\mu \Phi^\dagger {\cal D}^W_\mu \Phi) $, which
is the only one that contributes to the $\zeta_0$ kinetic term (while 
$ \tr({\cal D}^W_\mu U^\dagger {\cal D}^W_\mu U) $ does not). Upon approaching
the critical model, $(\rho, \eta) \to (\rho_{cr}, \eta_{cr})$, the canonically 
normalized $\zeta_0$ field has a squared mass $ m^2_{\zeta^0} \sim |\hat 
\mu_\Phi^2| / (1 - \bar \gamma) \to + \infty$. Decoupling of the $\zeta_0$
field in the critical model limit can easily be shown to imply
$\;|\tilde{C}| \leq O((1-\bar\gamma)^{1/2}) 
\stackrel{\rho \to \rho_{cr}, \eta \to \eta_{cr} }{\longrightarrow } 0 \;$  
for the coefficient $\tilde{C}$ of an otherwise expected term
$\; \tilde C \, \Lambda_S R \tr({\cal D}^W_\mu U^\dagger {\cal D}^W_\mu U)\;$
in $\Gamma_{4, cr}^{NG}$. \\
{\bf Remark V)}
The occurrence in $\Gamma_{4, cr}^{NG}$ of the NP terms $\propto C_1 \, , C_2$ 
is reflected in the presence of NP terms in the r.h.s.\ of the effective 
(renormalized) SDE associated to $\tilde\chi_{L,R} $ transformations, e.g.\ \\
\vspace*{-0.15cm}
\begin{equation}
Z_{\tilde J} \partial_\mu \tilde J^{L,\, i}_\mu = C_1 \Lambda_S
( \bar Q_L\frac{\tau^i}{2} U Q_R - {\rm h.c.} ) + C_2 \Lambda_S^2 i g_W 
\tr\Big( \Phi^\dagger [\frac{\tau^i}{2},W_\mu] {\cal D}_\mu^W \Phi + {\rm h.c.} \Big) \, .
\vspace*{-0.05cm}
\end{equation} 
Since the $\tilde\chi_{L,R}$ currents have 
vanishing anomalous dimension in the critical model, as it follows 
from the fact that for $\hat\mu^2_\phi >0$ they are conserved up 
to O($\Lambda_{UV}^{-2}$), the r.h.s.\ of the effective SDE
must be RGI in both the Wigner and the NG phase. In the latter case we
conclude that the operators $C_{1,Q} [\bar Q_L U Q_R + {\rm h.c.}]$ and
$C_2 \tr({\cal D}^W_\mu U^\dagger {\cal D}^W_\mu U)$ in $\Gamma_{4, cr}^{NG}$,
as well as their $\tilde\chi$-variations in the SDE, are RGI and UV-finite
-- just as it would happen for ordinary soft mass terms put by hand.
This implies that the dimensionless coefficients $C_{1,Q}$ and $C_2$ are
non-trivial functions of the various bare couplings (starting to O($g_S^4$)
or higher in $g_S$) and are endowed with a $\Lambda_{UV}$--dependence
that compensates for the one of $[\bar Q_L U Q_R + {\rm h.c.}]$ and 
$\tr({\cal D}^W_\mu U^\dagger {\cal D}^W_\mu U)$, respectively.

The model~(\ref{TOY+W}) is power counting renormalizable and the 
condition~(\ref{CRITPAR}) of maximal restoration of $\tilde\chi$ symmetries
fixes completely the parameters $\rho$ and $\eta$. If one were replacing
${\cal L}_{Wil}$ in the UV-regulated Lagrangian by some other set of 
$\tilde\chi$ violating $d > 4$ terms (respecting all the exact symmetries 
of ${\cal L}_{\rm toy+W}$), the condition of maximal $\tilde\chi$ symmetry
restoration would fix only $\eta_{cr}$ and one combination of the coefficients
in front the various $\tilde\chi$ violating $d > 4$ terms, but the low energy
physics would stay unchanged. Even in the presence of ``NP anomalies'' to
full $\tilde\chi$ symmetry restoration in the NG phase, the low energy physics
is expected to be controlled by the renormalized gauge couplings, the number 
of fermions and their transformation properties under the gauge groups. 
Universality and predictive power are in this sense preserved within the class of UV
regulated models endowed with the same exact symmetries as ${\cal L}_{\rm toy+W}$.
In the case of models with several types and/or generations of fermions, low energy
physics will in general also depend on the ratios of the critical parameters for 
the $\tilde\chi$ violating terms in the various fermion sectors. 

\vspace*{-0.1cm}
\section{Conclusions and outlook}

\vspace*{-0.1cm}
Based on the numerical evidence of Ref.~\cite{MGlat2018} in favour
of the elementary mass generation mechanism of Ref.~\cite{Frezzotti:2014wja},
we have discussed the key features of the model~(\ref{TOY+W}), which
in general carry over to possible extensions of phenomenological interest
and can be summarised as follows.\\
{\bf A)} The principle of maximal restoration of the (broken) $\tilde\chi_L \times 
\tilde\chi_R$ symmetry ensures both naturalness of
the effective masses and the appropriate form of universality
that holds at the NP level in $g_S^2$. \\
{\bf B)} The ``NP anomaly''  giving mass to the $Q$-fermions
is also responsible for the NP dynamical mass of weak gauge bosons,
which absorb the elementary NG-bosons present in the $g_W=0$ limit.

A lot remains to be done
to move towards more realistic theories with elementary particle
masses generated by ``NP anomalies'' as discussed above.
First, the model~(\ref{TOY+W}) should be
extended to three generations of quark and leptons, while the full  
EW interaction has to be included by promoting the exact
$\chi_L \times U(1)_Y$ invariance
to a gauge symmetry. Most importantly, a ``Tera-strong'' force and a set 
of ``Tera-fermions'' (name coming from Ref.~\cite{Glashow:2005jy}) that
communicate with ordinary matter (quarks or leptons) via strong and/or
EW interactions~\cite{FR18} should be also included in the model. 
The ``Tera-strong'' force is a new non-Abelian gauge interaction 
that becomes strong at the scale $\Lambda_T\gg\Lambda_{QCD}$, where $\Lambda_T$ 
can be roughly estimated to lie in the few TeV range~\cite{Frezzotti:2014wja,FR18},
if the experimental masses of the top quark and weak gauge bosons have to be
reproduced -- see Eq.~(\ref{W+Q+MASS}) with $\Lambda_S$ replaced by $\Lambda_T$.
%
As the condition of $\tilde\chi$ restoration entails the decoupling of the 
$\zeta_0$ component of the basic scalar field $\Phi$, one ends up with models of 
the Composite Higgs~\cite{PaniWulz} type, where the Higgs boson is
a bound state~\cite{Frezzotti:2014wja,FR18} in the $WW+ZZ$ channel 
that gets formed owing to the exchange of ``Tera-meson'' resonances
between two weak gauge bosons.

%

\bibliographystyle{JHEP}
\bibliography{bibliography.bib}

%

\end{document}